\documentclass[useAMS,usenatbib]{mn2e}

\usepackage{graphicx,amssymb,amstext}

\begin{document}

 \title[The IIFSCz]{The {\it Imperial IRAS-FSC Redshift Catalogue} (IIFSCz)} 
\author[L. Wang and M. Rowan-Robinson]{
\parbox[t]{\textwidth}{
Lingyu Wang$^{1}$\thanks{E-mail: lingyu.wang05@imperial.ac.uk}, Michael Rowan-Robinson$^{1}$}
\\
$^{1}$Astrophysics Group, Blackett Laboratory, Imperial College of Science Technology and Medicine, London SW7 2BZ, UK\\
}

\date{Accepted . Received ; in original form }

\maketitle

\begin{abstract}  
We present a new catalogue, the {\it Imperial IRAS-FSC Redshift Catalogue} (IIFSCz),  of 60,303 galaxies selected at 60\,$\mu$m from the {\it IRAS Faint Source Catalogue} (FSC). The IIFSCz consists of accurate position, optical, near-infrared and/or radio identifications, spectroscopic redshift (if available) or photometric redshift (if possible), predicted far-infrared (FIR) and submillimetre (submm) fluxes ranging from 12 to 1380\,$\mu$m based upon the best-fit infrared template. About $55\%$ of the galaxies in the IIFSCz have spectroscopic redshifts and a further $20\%$ have photometric redshifts obtained through either the training set or the template-fitting method. For S(60) $>$ 0.36 Jy, the 90$\%$ completeness limit of the FSC, 90$\%$ of the sources have either spectroscopic or photometric redshifts. Scientific applications of the IIFSCz include validation of current and forthcoming infrared and submm/mm surveys such as AKARI, \begin{em}Planck\end{em} and \begin{em}Herschel\end{em}, follow-up studies of rare source populations, large-scale structure and galaxy bias, local multiwavelength luminosity functions and source counts. The catalogue is publicly available from http://astro.imperial.ac.uk/$\sim$mrr/fss/.
\end{abstract}

\begin{keywords}
catalogues -- surveys -- galaxies: distances and redshifts --infrared: galaxies -- quasars: general -- large-scale structure of Universe.

\end{keywords}

\section{INTRODUCTION}
The {\it IRAS Faint Source Catalog} (FSC; Moshir et al. 1992) contains 173,044 sources extracted from image plates of co-added data. It is 2--2.5 times deeper than the {\it IRAS Point Source Catalogue} (PSC), reaching a depth of $\sim$0.2 Jy at 12, 25 and 60\,$\mu$m. The sky coverage of the FSC is limited to $|b|>20^\circ$ in unconfused regions at 60\,$\mu$m. For sources with high-quality flux density\footnote{In the IRAS FSC, the flux density quality (FQUAL) is classified as high (=3), moderate (=2) or upper limit (=1)}, the minimum reliability of the whole catalogue is $\sim99\%$ at 12 and 25\,$\mu$m and $\sim94\%$ at 60\,$\mu$m. Around 41$\%$ of the sources are detected at 60 $\mu$m (FQUAL = 2 or 3 at 60 $\mu$m).

The construction of a redshift catalogue of the FSC 60\,$\mu$m sources is made possible by overlaps (in terms of depth and area) with various surveys either spectroscopic or photometric, such as the Sloan Digital Sky Survey (SDSS; York et al. 2000), the Two Micron All Sky Survey (2MASS; Skrutskie et al. 1997) and the 6dF Galaxy Survey (Jones et al. 2004; Jones et al. 2005). We use two photometric redshift techniques, the empirical training set method and the Spectral Energy Distributions (SED) fitting procedure, to provide estimates of redshifts from optical and near-infrared (NIR) broad-band photometry. 

Other recent or planned all-sky surveys include AKARI and \begin{em}Planck\end{em}. The AKARI (previously known as ASTRO-F) All-Sky Survey, which ended on August 26th, 2007, has observed $94\%$ of the sky from mid- to far-infrared (Murakami et al. 2007). The $5\sigma$ point source detection limit of the Far-Infrared Surveyor with a single scan coverage is estimated to be 2.4, 0.55, 1.4 and 6.3 Jy at 65, 90, 140 and 160 micron respectively (Kawada et al. 2007; Wang et al. 2008). The ESA (European Space Agency) mission \begin{em}Planck\end{em} is going to map the Cosmic Microwave Background (CMB) with unprecedented angular resolution and sensitivity as well as produce all-sky catalogues of infrared and radio galaxies in the frequency bands ranging from 30 to 850 GHz. The study of these \begin{em}Planck\end{em} extragalactic point sources can not only help clean up the foreground contaminants in the CMB images but also constrain galaxy formation and evolution models and allow searches for high redshift dusty galaxies. For the HFI  channels, the estimated \begin{em}Planck\end{em} All Sky Survey sensitivity at the $3\sigma$ level is 26, 37, 75, 180, 300 mJy at 2100, 1380, 850, 550, 350\,$\mu$m respectively (\begin{em}Planck\end{em} Bluebook, ESA-SCI(2005)-1, Version 2) and thus many of the sources detected by \begin{em}Planck\end{em} will also be present in the IRAS catalogues. Another ESA mission \begin{em}Herschel\end{em} (Pilbratt 2004), which is to be launched together with \begin{em}Planck\end{em}, will observe the universe in the far-infrared (FIR) and submillimetre (submm) range (approximately 57 -- 670\,$\mu$m) with the formation and evolution of galaxies and stars and stellar systems as its key science goals. The SPIRE instrument (Griffin et al. 2007) will perform imaging in the broadband photometry mode centred at 250, 350 and 500\,$\mu$m with the predicted point source sensitivity in the range 8 -- 11 mJy ($5\sigma$, 1 hr). To aid these missions, we have provided predicted fluxes at the relevant mission wavelengths. The Imperial IRAS-FSC Redshift Catalogue (IIFSCz) will be an important input and validation catalogue for current and forthcoming wide-area infrared/submm surveys described above. 

There are also several extragalactic science programmes that can be carried out with the IIFSCz. The convergence of the cosmological dipole is particularly worth further investigation as previous measurements with all-sky surveys such as IRAS or 2MASS have not yet shown a consensus on the convergence depth (Rowan-Robinson et al. 2000; Maller et al. 2003; Erdo\v{g}du et al. 2006). While 2MASS samples the local universe (within 200 $h^{-1}$ Mpc) very well, it is not quite deep enough to detect potential contributions to the dipole signal from large distance. The IIFSCz provides a huge sample for large-scale structure and velocity studies, e.g. the baryon acoustic oscillation (Eisenstein et al. 2005), the two-point correlation function  and its dependence on the FIR luminosity and star formation rate (Mann, Saunders \& Taylor 1996; Szapudi et al. 2000; Hawkins et al. 2001), the local multiwavelength luminosity functions (Serjeant \& Harrison 2005), follow-up studies of rare source populations such as ultraluminous infrared galaxies (ULIRGs) and hyperluminous infrared galaxies (HLIRGs) and number counts. 

The layout of this paper is as follows. The criteria used to select galaxies from the IRAS FSC are described in Section~\ref{sec:selection}. The issue of extended sources is discussed in Section~\ref{sec:ExtendedSources}. In Section~\ref{sec:identification}, we firstly obtain spectroscopic redshifts from a number of databases and then cross-identify (using the likelihood ratio technique) FSC sources with their optical, near-infrared and/or radio counterparts. In Section~\ref{sec:ANNz}, we carry out photometric redshift estimation (using both the training set and the template-fitting method) and then make flux predictions at FIR and submm wavelengths. The overall properties of the IIFSCz is described in Section~\ref{sec:catalogueDescriptions}. Finally, discussions and conclusions are given in Section~\ref{sec:discussions} and Section~\ref{sec:conclusions} respectively. We adopt a flat cosmological model with $\Lambda=0.7$ and $h_0=0.72$.

\begin{figure}
\includegraphics[height=2.8in,width=3.4in]{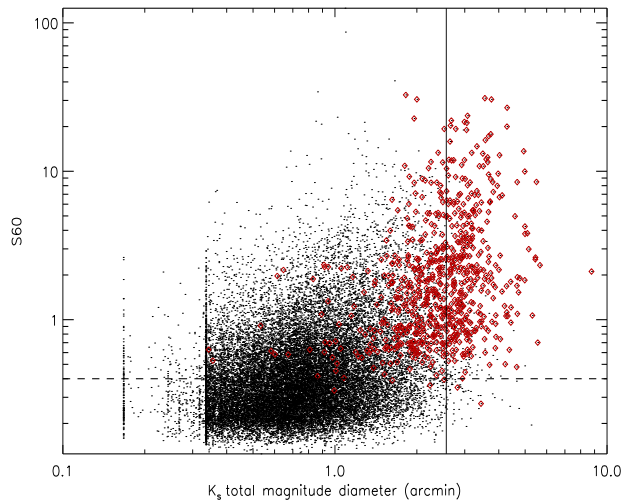}
\caption{Flux densities at 60\,$\mu$m vs. $K_s$-band total magnitude diameter for around 28,000 FSC sources with 2MASS identifications. The red symbols are the FSC sources associated with extended PSCz sources (i.e. $D_{25}>2.25'$). The solid line shows the mean $K_s$ total magnitude diameter $D_{Ktot}=2.57'$ for the red symbols and the dashed line is where $S60=0.4$ Jy.}
\label{fig:flux-size}
\end{figure}

\begin{figure}
\includegraphics[height=2.95in,width=3.4in]{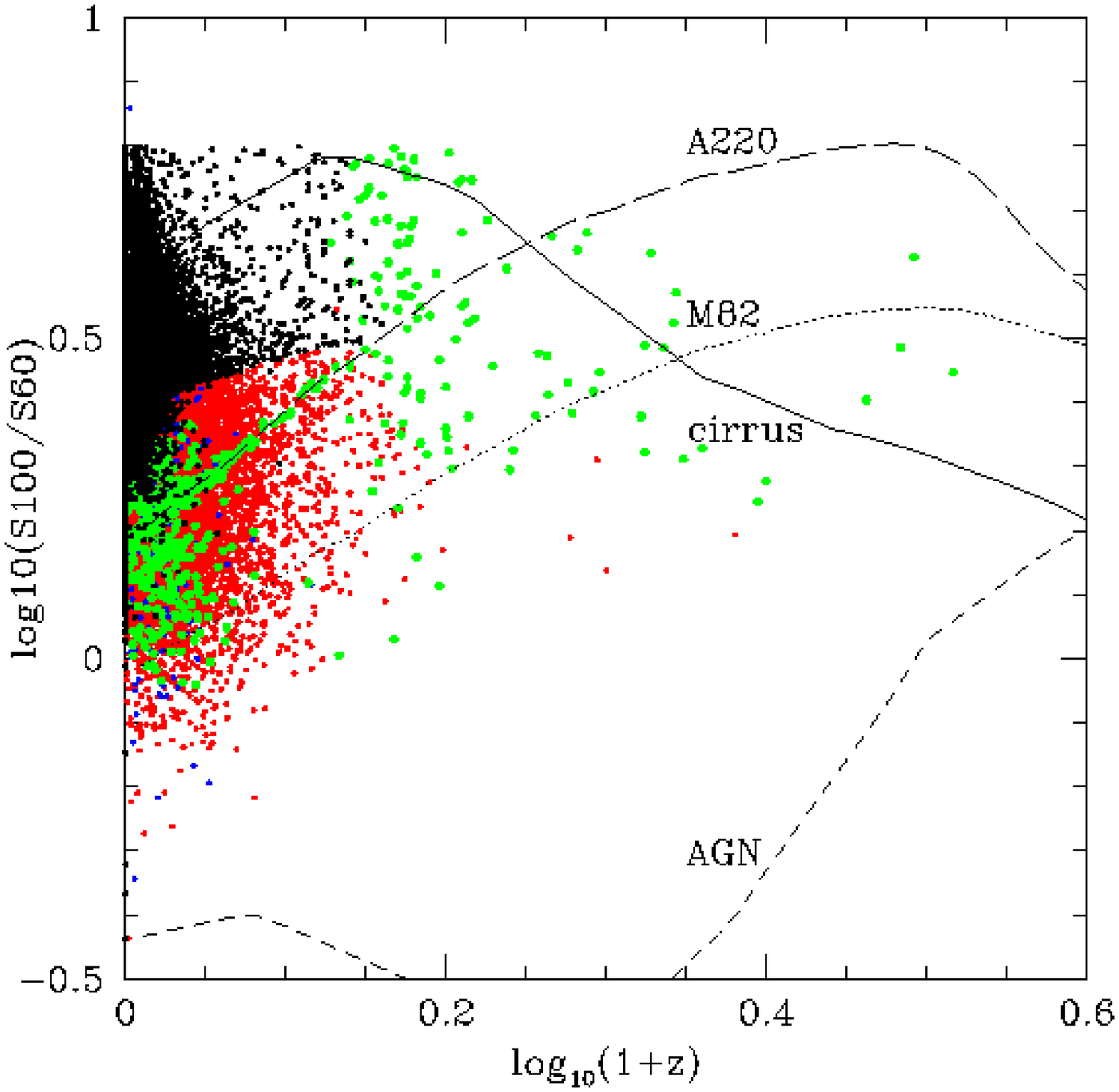}
\includegraphics[height=2.95in,width=3.4in]{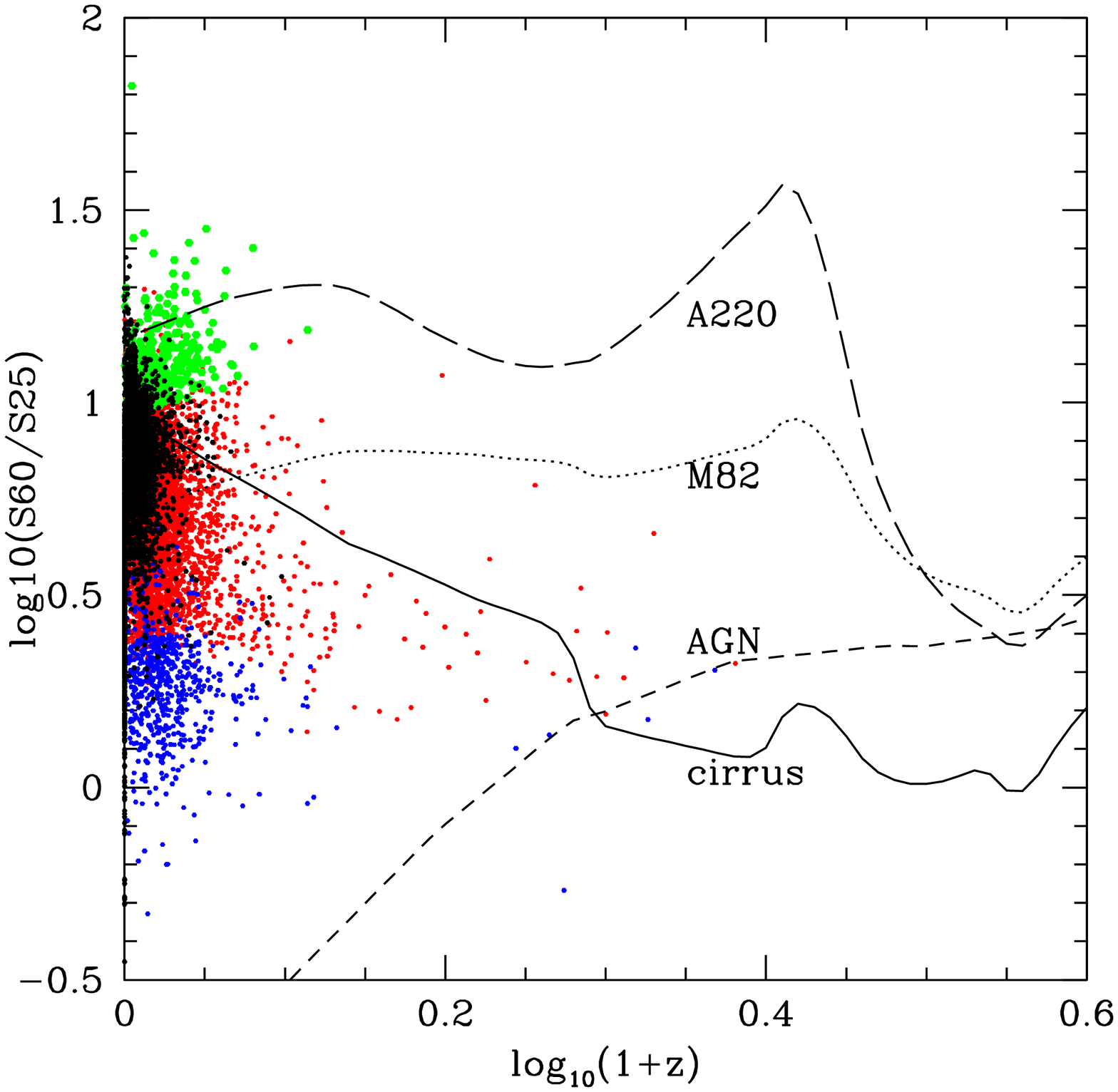}
\includegraphics[height=2.95in,width=3.4in]{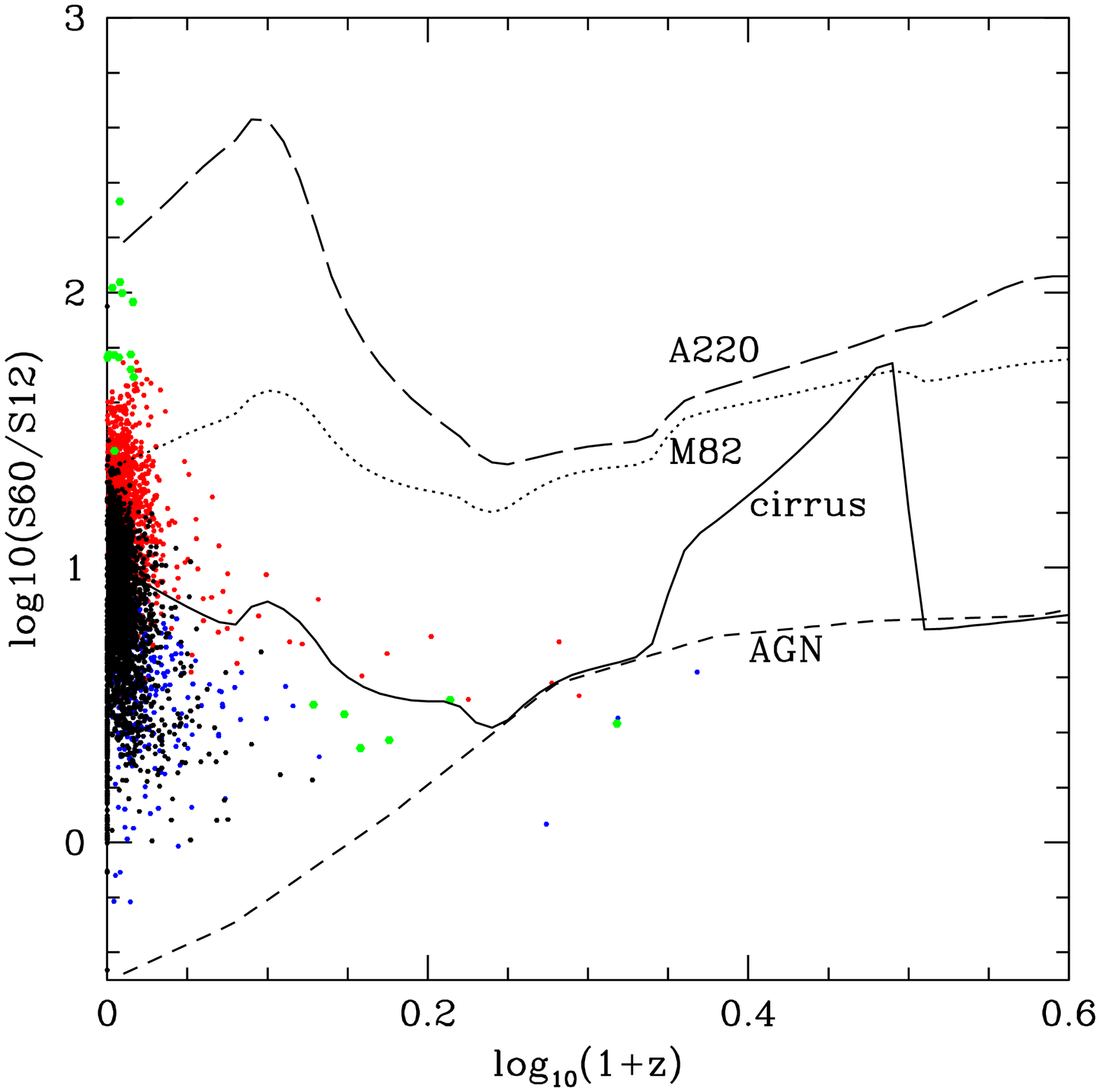}
\caption{Upper: Colour at 100-60\,$\mu$m vs. spectroscopic redshift colour-coded by four infrared templates (black: cirrus; red: M82; green: A220; blue: AGN dust torus). Middle: Colour at 25-60\,$\mu$m vs. spectroscopic redshift. Lower: Colour at 12-60\,$\mu$m vs. spectroscopic redshift.}
\label{fig:colour-z}
\end{figure}

\section{CATALOGUE CONSTRUCTION}

\subsection{Sample selection}
\label{sec:selection}
To obtain a complete sample of galaxies from the IRAS FSC, our selection criteria are:
 
(1) To ensure reliability, we select sources with FQUAL $\geq3$ and SNR $>5$ at 60\,$\mu$m. There are two types of signal-to-noise ratio in the IRAS FSC, the SNR at a given pixel (LOCSNR) and the SNR which uses a noise value derived from a local region. The latter is adopted here. This condition leaves us with 63,842 sources;

(2) To exclude cirrus, we require $\log$(S100/S60)$<0.8$ if FQUAL $\geq2$ at 100\,$\mu$m. The sample size is reduced to 63,117. The upper panel in Fig.~\ref{fig:colour-z} shows the predicted 100-60\,$\mu$m colour as a function of redshift based upon four infrared templates, cirrus, M82 starburst, Arp 220 starburst and AGN dust torus (Rowan-Robinson et al. 2004; Rowan-Robinson et al. 2008). It is clear that our constraint on the 60-100\,$\mu$m flux ratio should not exclude any infrared galaxy type for $z<4$; 

\begin{figure*}
\includegraphics[height=3.3in,width=4.8in]{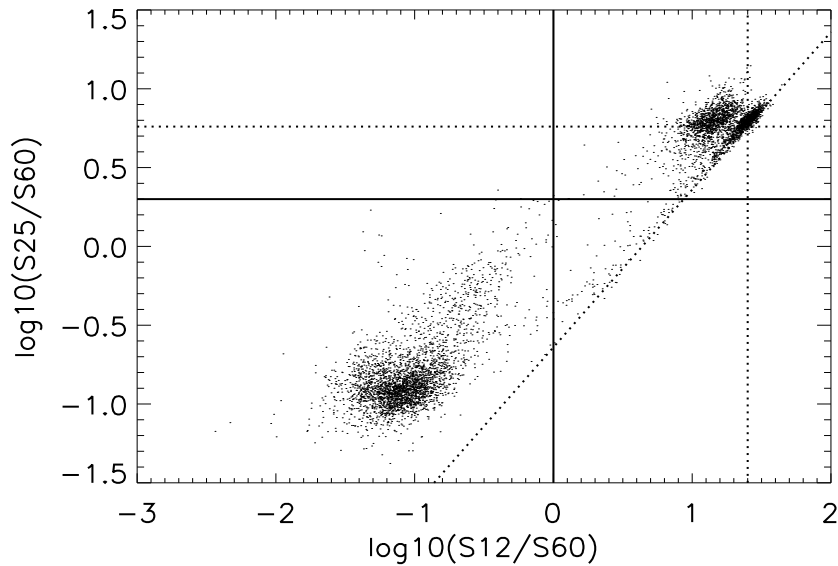}
\caption{Colour-colour distribution of sources detected at 12 and 25\,$\mu$m after applying the cirrus rejection criterion. The dotted lines are the Rayleigh-Jeans predictions, while the solid lines indicate our stellar rejection criteria. Sources with $\log10$(S25/S60)$<0.3$ and $\log10$(S12/S60)$<0$ are selected as galaxies.}
\label{fig:12-25-60}
\end{figure*}

(3) To discriminate against stars, we firstly require $\log$(S60/S25)$>-0.3$ if FQUAL $\geq2$ at 25\,$\mu$m and then $\log$(S60/S12)$>0$ if FQUAL $\geq$2 at 12\,$\mu$m. A total of 60,381 sources have met the above criteria, the faintest of which has a flux density of 0.12 Jy at 60\,$\mu$m. Fig.~\ref{fig:12-25-60} is the colour-colour diagram of FSC sources with detections at 12 and 25\,$\mu$m after applying the 60-100\,$\mu$m colour cut. The Rayleigh-Jeans predictions for the 12-60, 25-60, 25-12\,$\mu$m flux ratio are the dotted lines, while our stellar rejection criteria are indicated by the solid lines. The concentrations of objects to the upper right are stars. Admittedly, our colour cuts might be a little harsh and therefore some low-redshift AGNs might be missing from our sample (see the middle and lower panel in Fig.\ref{fig:colour-z}). By examining DSS (Digitized Sky Survey) images and cross-identification in NED or Simbad (if available) of sources in regions where $0<\log$(S12/S60)$<0.7$ and $\log$(S25/S60)$>0$ or $0<\log$(S12/S60)$<0.3$ and $\log$(S25/S60)$<0.3$, we managed to retrieve 3 Seyfert galaxies, 8 unidentified or confused galaxies, while the rest are mostly carbon stars, post-AGB stars etc. At this stage, our sample contains $60,392$ galaxies which forms the base catalogue of the IIFSCz. However, we point out that the sample size will undergo one more change in Section~\ref{sec:redshiftCompilation}.

\subsection{Extended sources}
\label{sec:ExtendedSources}
In the IRAS Faint Source Survey, the median filtering method was used to maximise detection of faint point sources. However, the fluxes of extended sources are attenuated. This may cause nearby galaxies to be incorrectly excluded from our catalogue. In the construction of the PSCz catalogue, Saunders et al. (2000) used the ADDSCAN/SCANPI to derive fluxes for sources with blue-light isophotal major diameters $D_{25}>2.25'$. For sources with $D_{25}>8'$, they used fluxes from the catalogue of IRAS observations of 85 large optical galaxies (Rice et al. 1988). There are a total of 1402 galaxies with blue isophotal diameters $>2.25'$ in the PSCz, 1290 of which are associated with PSC sources. Of the 1290 PSC sources, 938 were found in the FSC after applying our first selection criterion and so we have adopted their PSCz fluxes. Of the remaining 112 sources not found in the PSC, we identified 76 in the FSC and their fluxes were also replaced by PSCz fluxes. Some of the extended PSCz sources were not found in the FSC because they lie at $|b|<20^\circ$.

ADDSCAN was designed for accurate flux measurement of bright sources and extended sources which need higher in-scan resolution. From a sample of 62 unambiguously extended galaxies in the Virgo Cluster, the ratio of the addscan flux to the FSC flux stays close to unity for sources fainter than $\sim$0.4 Jy at 60\,$\mu$m (Moshir et al. 1992). Indeed, almost all FSC sources cross-matched with extended PSCz sources ($D_{25}>2.25'$) have $S60>0.4$ Jy. In order to estimate the likely number of FSC sources to have their fluxes seriously underestimated, we show the $K_s$-band total magnitude diameter ($D_{Ktot}$) for around 28,000 FSC sources with 2MASS identifications (see Section 3.3) in Fig.~\ref{fig:flux-size}. The mean $K_s$ total magnitude diameter for FSC sources associated with extended PSCz sources (red symbols in Fig.~\ref{fig:flux-size}) is $D_{Ktot}=2.57'\pm0.90'$. Therefore, the $K_s$ total magnitude diameter $D_{Ktot}$ is approximately the same as the blue isophotal diameter $D_{25}$ (Jarrett et al. 2003). In total, there are 1677, 430, 97 galaxies with $S60>0.4$ Jy and $D_{Ktot}>1.7', 2.6', 3.5'$ respectively. Excluding extended sources whose fluxes have already been replaced by PSCz fluxes, we conclude there is still $<\sim1\%$ of the galaxies in our catalogue suffer from flux underestimation.

\section{SOURCE IDENTIFICATION}
\label{sec:identification}

\subsection{Redshift compilation from NED, FSSz, PSCz \& 6dF}
\label{sec:redshiftCompilation}
Having constructed the base catalogue, the next step is to obtain spectroscopic redshifts from past redshift surveys and the literature. The NASA/IPAC Extragalactic Database (NED) currently contains 10.4 million objects, 1.4 million redshifts and 16.0 million cross-identifications based upon astrometry, photometry and avoidance of confusion. 

\begin{figure}
\includegraphics[height=2.8in,width=3.4in]{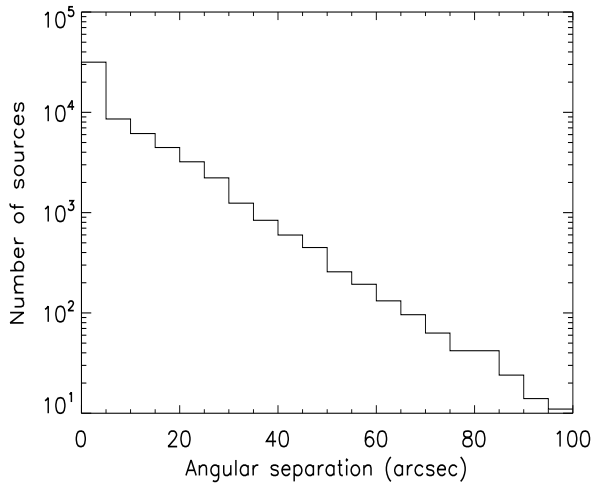}
\caption{Distribution of the angular separations between the IRAS positions and the NED positions of the FSC sources.}
\label{fig:crossmatch_hist}
\end{figure}

An NED all-sky query\footnote{An all-sky query in NED usually returns a list of sources with information such as source name, type, position and redshift. However, photometric data for each source is not included and has to be retrieved individually.} of FSC sources at $|b|>20^\circ$ with S60$>$0.1 Jy returned 64,219 objects (we will refer to this sample as NED-FSC). In the NED, two constraints were used to select FSC sources, $\log$(S60/S25)$>-0.3$ and FQUAL$\ge 3$ at 60\,$\mu$m. Therefore, our sample is expected to be smaller than the NED-FSC. The NED position of a given FSC source is the best possible position, that is to say optical or near-infrared position if available and IRAS position if not. The accuracy of IRAS position depends on the size, brightness and spectral energy distribution of the source but is usually $<$20" (1-$\sigma$). Within a radius of 300", 60,320 out of 60,392 sources in our base catalogue are matched with sources in the NED-FSC, more than 23,000 of which have spectroscopic redshifts in the NED. A manual checking of the matched sources with separations $\ge$80" ($\sim$170 objects) proved that these cross-identifications are correct. Fig.~\ref{fig:crossmatch_hist} shows the number of cross-identifications as a function of the angular separation between the IRAS position and the NED position. 3 FSC sources are matched manually to their cross-identifications due to their large positional separations. 44 FSC sources are not found in NED. In addition, we have removed a few FSC sources which are identified in NED as galactic objects from our base catalogue. This is the final change to the size of our base catalogue which now contains 60,303 galaxies. 

The FSS redshift survey (FSSz; Oliver, PhD thesis) covering 700 deg$^2$ measured redshifts for 1,546 FSC sources. It provides an additional 568 redshifts for our catalogue.

So far, around 2,000 bright FSC sources (S60$\geq$0.6 Jy) in our base catalogue do not have spectroscopic redshifts. The PSC Redshift Survey (PSCz; Saunders et al. 2000) is complete to S60=0.6 Jy. NED has not folded in the PSCz due to issues such as significant positional difference and incorrect NGC/IC/MCG cross-identifications. We managed to get 1,972 spectroscopic redshifts from the PSCz via exact source name matching. 

The 6dF Galaxy Survey covers roughly 2/3 of the southern sky, measuring redshifts for over 80,000 galaxies. The primary redshift sample is selected from the 2MASS Extended Source Catalog (XSC), together with 13 other samples such as the SuperCOSMOS catalogue, the ROSAT All-Sky Survey and the IRAS FSC. NED used a 3" radius in matching the 6dF with the FSC, however, matches were not made if there were more than one 6dF object within the IRAS beam. In total, NED has matched $\sim6000$ FSC sources with the 6dF observations.

In the 6dF database, 10111 objects appear in both the IRAS FSC and the SPECTRA table where all the observational and redshift related information are stored. Again, by matching FSC source names, the 6dF provides an extra 2,794 redshifts with acceptable quality (on quality scale Q=3 or 4).

To summarise, we have obtained 29,037 spectroscopic redshifts from NED, FSSz, PSCz and 6dF, which comprises $48\%$ of our base catalogue. However, there are a few thousand more redshifts to be gained from the SDSS DR6 survey (see below).

\subsection{CROSS-MATCH WITH SDSS}
The SDSS photometric DR6 survey has covered 9,583 deg$^2$ at $u, g, r, i, z$ with magnitude limits 22.0, 22.2, 22.2, 21.3, 20.5 respectively. The spectroscopic DR6 survey has covered 7,425 deg$^2$, the main samples of which are magnitude-limited at Petrosian r$<$17.77 for galaxies. 

As mentioned in Section~\ref{sec:redshiftCompilation}, one does not get photometric data using the NED all-sky search. Therefore, in order to get optical magnitudes, we have to cross-identify FSC sources with their optical counterparts in the SDSS DR6 catalogues. In addition, because NED has only entered galaxies from the spectroscopic DR5 survey (covering 5,740 deg$^2$) so far, we can also get new redshifts from the spectroscopic DR6 survey. 

\subsubsection{The likelihood ratio technique}
We use the likelihood ratio technique (LR; Wolstencroft et al. 1986; Sutherland \& Saunders 1992; Ciliegi et al. 2003; Brusa et al. 2007) to cross-identify FSC sources with their optical counterparts. The LR essentially compares the probability of a candidate being the true counterpart as a function of magnitude $m$ and separation $d$ with that of a chance association, i.e. 
\begin{equation}
L=\frac{p(m,d)~dm~dx~dy}{n(m)~dm~dx~dy}=\frac{q(m)f(d)}{n(m)}.
\end{equation}
The probability distribution function $f(d)$ is usually assumed to be a two-dimensional Gaussian, 
\begin{eqnarray}
f(d) & = & \frac{1}{\sigma_1 \sigma_2 2\pi} \exp[-\frac{1}{2}(\frac{d_1^2}{\sigma_1^2}+\frac{d_2^2}{\sigma_2^2})] \nonumber \\
     & = & \frac{1}{\sigma_1 \sigma_2 2\pi} \exp(-\frac{1}{2} d^2), 
\end{eqnarray}
where $\sigma_1$ and $\sigma_2$ are the axes of the positional uncertainty ellipse of a given FSC source, $d_1$ and $d_2$ are the positional separation along the axes and $d$ is the normalised angular distance. For FSC sources, the mean 1-$\sigma$ positional uncertainty along the minor and major axes are 5 and 18 arcsec respectively. The magnitude distribution function of background objects $n(m)$ is taken from objects around some random positions in the sky. The magnitude distribution function of the true optical counterparts $q(m)$ is obtained by subtracting $n(m)$ from the magnitude distribution of objects each FSC sources. In Bayesian inference, it can be shown that $L$ contains all the information about the probability of a candidate being the true counterpart. In cases where there are multiple candidates, the one with the highest likelihood is selected as the true counterpart.

\subsubsection{Obtaining redshifts from SDSS DR6}
\label{Sec:DR6redshifts}
Firstly, we need to test the reliability of the LR method. This is easily achievable as we have already obtained 29,022 spectroscopic redshifts in Section~\ref{sec:redshiftCompilation} which can be used to compare with the redshifts of the optical counterparts found in the spectroscopic DR6 survey. 

In the spectroscopic DR6 catalogue, we search for all primary objects\footnote{Whenever the SDSS makes multiple observations of the same object, the one with the best photometry will be assigned as the `primary' observation.} within 1 arcmin from the IRAS position of each FSC source with known spectroscopic redshift gained in Section~\ref{sec:redshiftCompilation}. Although in principle we can use accurate positions for these FSC sources, the IRAS positions are used to investigate the impact of large positional uncertainty. 

\begin{figure*}
\includegraphics[height=3.5in,width=4.6in]{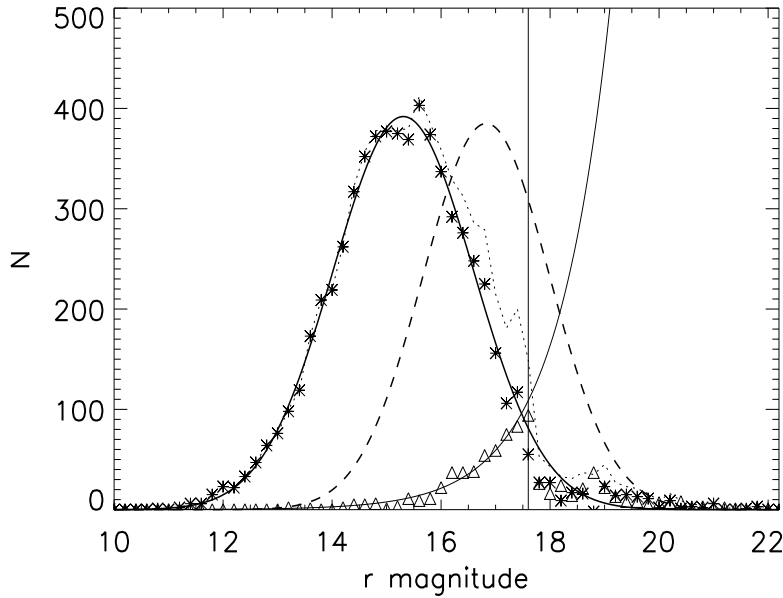}
\caption{The r magnitude distribution of objects in the SDSS spectroscopic DR6 around FSC sources with redshifts obtained in Section~\ref{sec:redshiftCompilation} (black dotted line). The triangles show the magnitude distribution of random background objects, i.e. $n(m)$, and the thin solid curve is an exponential fit to these points. The asterisks depict the magnitude distribution of the true optical counterparts, i.e. $q(m)$, and the thick solid line is a Gaussian fit to these points. The dashed curve is a Gaussian fit to $q(m)$ for optical objects around FSC sources which do not gain redshifts in Section~\ref{sec:redshiftCompilation}. The vertical line is where r=17.6.}
\label{fig:q_and_n}
\end{figure*}

Of the 29,022 FSC sources with spectroscopic redshifts, 8,979 were found to have optical counterparts and, of these, an average of 1.2 optical counterparts were found for each. We exclude candidates with $r>17.6$ (above the magnitude limit) and $z<0.0005$ in order to exclude stars. Fig. \ref{fig:q_and_n} shows the r-band magnitude distribution of the true counterparts and that of the random background objects. The thick solid line is a Gaussian fit to $q(m)$, 
\begin{equation}
q(m) \propto \exp[ -\frac{1}{2} (\frac{m-15.31}{1.29})^2 ]
\end{equation}
and the thin solid curve is an exponential model of $n(m)$. Using the LR method, we managed to cross-match 8166 FSC sources with optical galaxies in the spectroscopic DR6 survey. For an FSC source, if its redshift obtained in Section~\ref{sec:redshiftCompilation} agrees with the redshift of its optical counterpart, then we say that the cross-identification is correct. Thus, the reliability, defined as the ratio of the number of correction identifications to the total number of sources in the cross-matched sample, 
\begin{equation}
Reliability=\frac{N_{correct}}{N_{total}},
\end{equation}
of the cross-matched sample is estimated to be $\sim97\%$.

Now we can apply the same procedure to the 31,260 FSC sources which did not receive spectroscopic redshifts in Section~\ref{sec:redshiftCompilation}. Using a search radius of 1', 4654 FSC sources are matched with 5462 optical candidates. The dashed curve in Fig.~\ref{fig:q_and_n} is a Gaussian fit to the new $q(m)$, where the peak is now shifted to $\sim$16.8. The LR technique gives rise to new optical identifications and spectroscopic redshifts for 3,844 FSC sources.

\subsubsection{Obtaining photometry from SDSS DR6}

Source identification becomes much more complicated in matching FSC sources with the photometric DR6 catalogue as the probability of chance association is $\propto n \pi d^2$, where n is the number density of background optical objects. Using a search radius of 1', an average of 28 optical counterpart candidates were found for each FSC source. The LR approach produced 9,425 cross-identifications between the sample of 31,260 FSC sources and the photometric DR6 catalogue. The adopted $q(m)$ is the blue dotted line in Fig.~\ref{fig:q_and_n}. SDSS objects classified as stellar are rejected not only because the contamination caused by stars is significant but also because the vast majority of the QSOs already have redshifts. 

To estimate the reliability of our cross-matched sample, firstly let $\mathbb{F}$ denote the set of 9,425 FSC sources and $\mathbb{O}$ the set of optical counterparts. The reliability of $\mathbb{O}$ is simply the ratio of the number of true optical ids in $\mathbb{O}$ to the size of $\mathbb{O}$. Consider a subset $\mathcal{F} \subset \mathbb{F}$ for which the true optical counterparts $\mathcal{T}$ are known. If the set of optical counterparts of $\mathcal{F}$ found using the LR method is denoted $\mathcal{O} \subset \mathbb{O}$, the reliability of $\mathcal{O}$ is given by the ratio
\begin{equation}
Reliability = \frac{ N_{ \mathcal{O} \cap \mathcal{T} } }{ N_{\mathcal{O}} },
\end{equation}
where the numerator is the size of the intersection between $\mathcal{O}$ and $\mathcal{T}$ and the denominator is the size of $\mathcal{O}$. Assuming the subset $\mathcal{F}$ is representative of $\mathbb{F}$, the reliability of the subset $\mathcal{O}$ should reflect the reliability of $\mathbb{O}$.

In order to get reliable optical identifications, accurate positions of the FSC sources are needed so as to reduce the probability of chance association. In the following, three test samples are used: 

\begin{itemize}
\item We select 5,000 FSC sources with 2MASS identifications, the positional uncertainty of which is around 1.25" at the $95\%$ confidence level. Using a search radius of 1" around the 2MASS positions, 1,325 FSC sources (out of 5,000) were matched with unique optical counterparts in the photometric DR6 survey (we will refer to this set of optical ids as `2MASS\_1arcsec'). Using `2MASS\_1arcsec' as $\mathcal{T}$, the reliability of $\mathbb{O}$ is estimated to be $83\%$. Excluding `stellar' optical objects, the reliability is increased to $88\%$.

\item The Faint Images of the Radio Sky at Twenty centimetres survey (FIRST; Becker et al. 2004) with a sensitivity of 1 mJy at 1.4 GHz and an angular resolution of 5" yielded a catalogue of $\sim$811,000 sources. The survey area overlaps with that of the SDSS and the radius of the $90\%$ confidence error circle is less than 1". An average of 1.1 FIRST sources were found within 1' from an FSC source. We have selected 4,886 unambiguous FSC-FIRST cross-matches, 3,273 of which have unique optical counterparts in the photometric DR6 survey (`FSC-FIRST-SDSS'). Using `FSC-FIRST-SDSS' as $\mathcal{T}$, the reliability of $\mathbb{O}$ is $80\%$.

\item The sample of 3844 new optical identifications obtained in Section~\ref{Sec:DR6redshifts} serves as the third test sample and it gives us a reliability of $76\%$. 
\end{itemize}

Thus, on average the optical identifications obtained from the SDSS photometric DR6 are $\sim80\%$ reliable.

\subsection{CROSS-MATCH WITH 2MASS}
The 2MASS has uniformly surveyed the whole sky in $J$, $H$ and $K_s$, reaching a median depth of $z=0.073$. To date, the 2MASS XSC has been comprehensively assimilated and cross-matched with other surveys in NED. Therefore, it is fairly straightforward to retrieve 28,640 cross-ids between the IIFSCz and the 2MASS XSC, around 7,000 of which do not have redshift information. We have also identified 39 FSC sources in the 2MASS Point Source Catalogue (PSC), 25 of which are shown to be QSOs. 

\subsection{CROSS-MATCH WITH NVSS}
The NRAO VLA Sky Survey (NVSS; Condon et al. 1998) is a moderately deep radio survey over $82\%$ of the sky. It has produced a catalogue of nearly 2 million sources brighter than 2.5 mJy at 1.4 GHz. The observed extragalactic sources include nearby normal galaxies, AGNs, star-forming galaxies, starbursts etc. The rms positional uncertainties are $\leq$ 1" for sources brighter than 15 mJy and 7" for sources above the detection limit. Given the almost linear FIR to radio luminosity correlation (Helou, Soifer \& Rowan-Robinson 1985; de Jong et al. 1985; Yun, Reddy \& Condon 2001; Appleton et al. 2004), the NVSS is expected to detect most of the FSC sources and therefore provide much more accurate positions for these sources. 

Unique radio counterparts were found for 23,183 FSC sources within a radius of 1' and multiple counterparts were found in $\sim$440 cases. Thus the cross-identification accuracy is $\ge98\%$ even if we adopt a crude nearest-object cross-matching approach. The NVSS provides cross-ids for 3123 FSC sources without any observation other than IRAS. 

\section{PHOTOMETRIC REDSHIFT ESTIMATION}
\label{sec:ANNz}

\subsection{THE TRAINING SET METHOD}
We use two different techniques, the empirical training set and the template-fitting technique, to estimate photometric redshifts for FSC sources with optical, NIR and/or radio photometry. In this section, we apply the public Artificial Neural Networks code of Firth, Lahav \& Somerville 2003; Collister \& Lahav 2004 ($ANNz$) to the IIFSCz and results are presented and discussed. 

$ANNz$ requires a representative training set\footnote{A training set is a collection of sources with accurate redshifts and the same filter set as the testing set for which we would like to estimate the photometric redshifts. A Validation set is a random collection of sources with known redshifts. However, the validation set is not used in the network training process.} to learn the functional relationship between photometry and redshift. Advantages of the training set method include nullifying systematic effects, freedom and flexibility in choosing input parameters, greater accuracy and efficiency etc. Having said that, the photometric redshift accuracy strongly depends on the quality of the training set. In principle, the training set should occupy the same region in the parameter space (colour, redshift, spectral type etc.) as the testing set. 

The structure of $ANNz$ can be roughly divided into three passages, the input layer, the intervening/hidden layers and the output layer. In our case, it can be viewed as: magnitudes at various wavebands $\rightarrow$ functional mapping $\rightarrow$ photometric redshift. Each layer consists of a number of nodes. The network architechture is determined by the number of filters at the input layer, the number and size of each of the hidden layers and the number of outputs. For example, a 3.10.10.10.1 network architecture has photomeric measurements in three filters, three hidden layers each of which has ten nodes, and one output, i.e. the derived photometric redshift. The optimisation of the mapping is achieved by tuning the network weights associated with connected nodes. The training process is terminated when the cost function, defined as $(z_{phot} - z_{spec})^2$, is minimal on the validation set. $ANNz$ gives two types of errors, photometric noise and network variance. The latter is obtained by using a number of networks known as a committee.

\begin{figure}
\includegraphics[height=2.8in,width=3.4in]{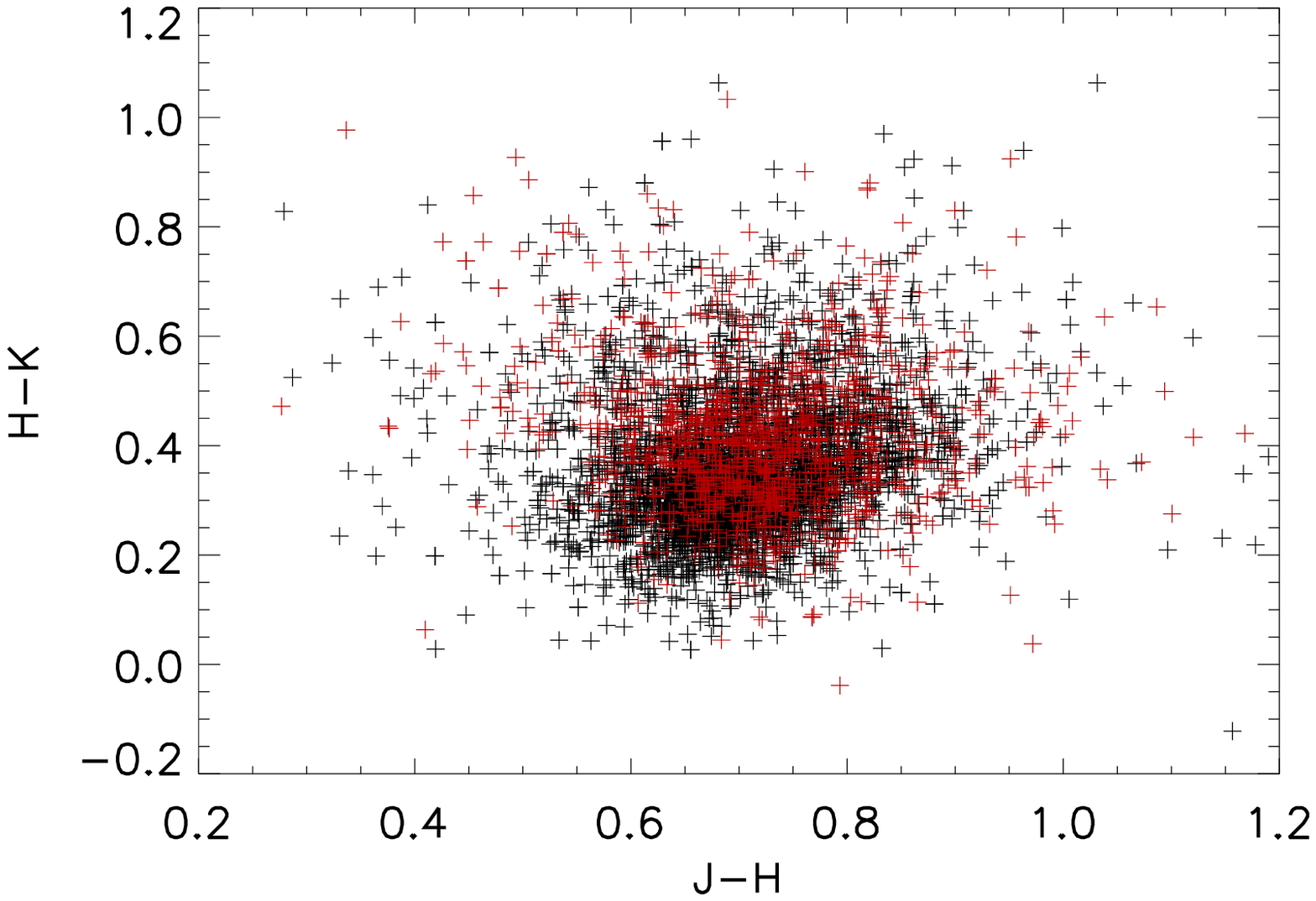}
\includegraphics[height=2.8in,width=3.4in]{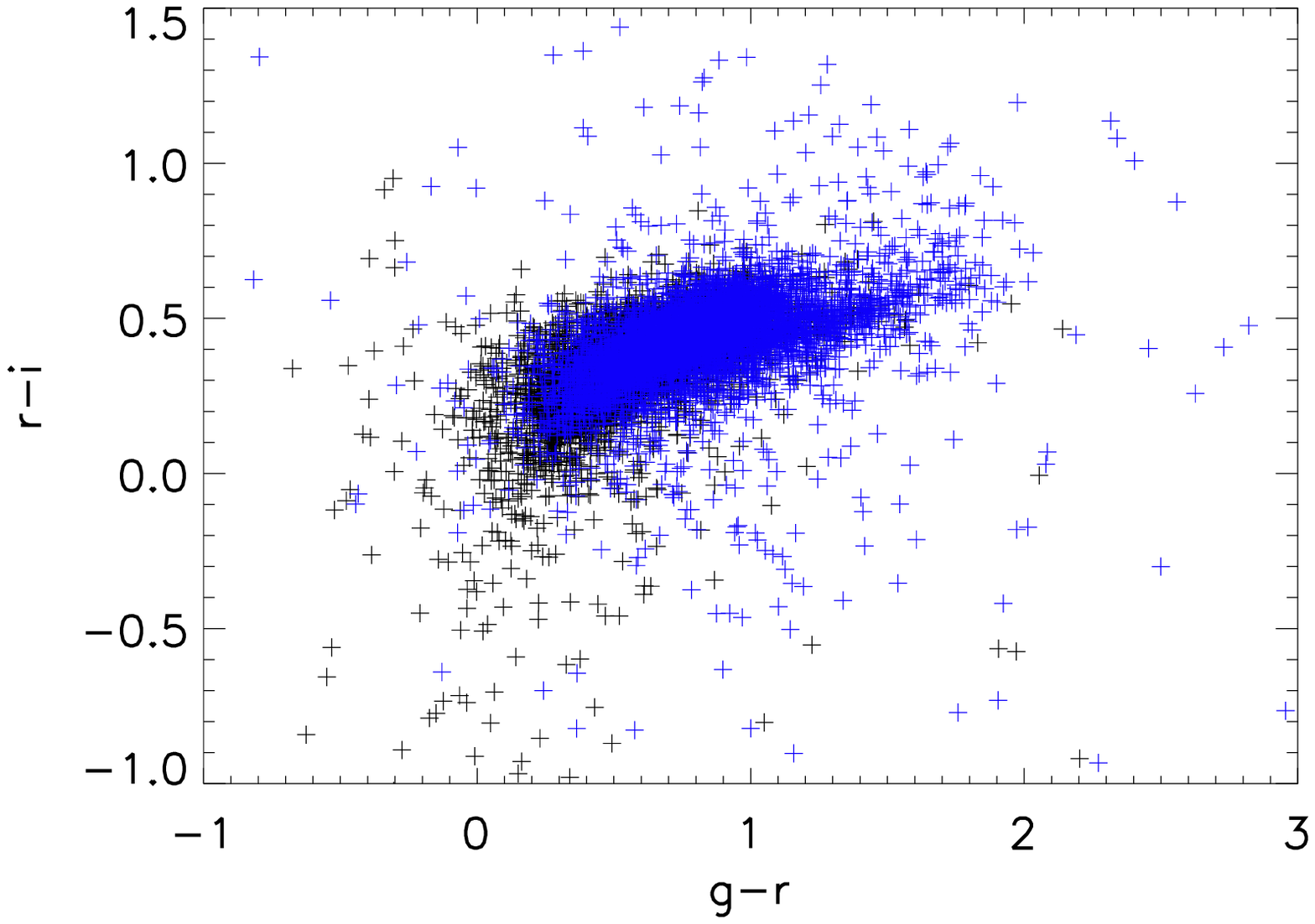}
\caption{Colour-colour distribution of FSC sources with 2MASS (upper) or SDSS cross-ids (lower).}
\label{fig:colour_dist}
\end{figure}

\begin{figure}
\includegraphics[height=2.8in,width=3.4in]{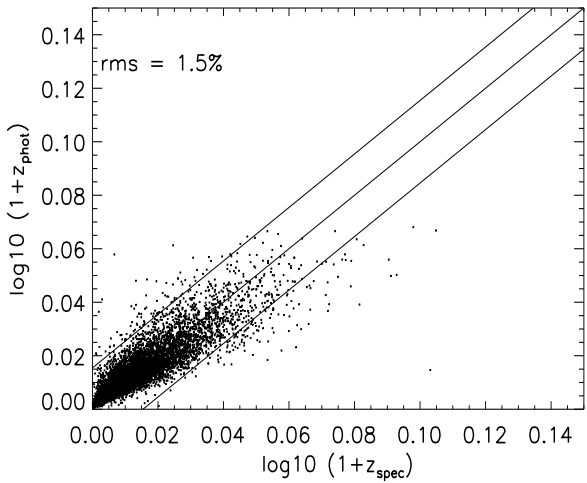}
\includegraphics[height=2.8in,width=3.4in]{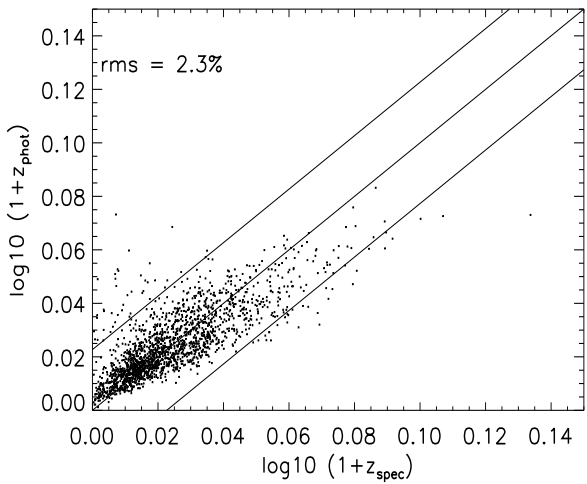}
\caption{Upper: Photometric redshift versus spectroscopic redshift, using 3 near-infrared bands and a 3.10.10.10.1 network architecture. Lower: Photometric redshift versus spectroscopic redshift, using 5 optical bands and a 5.10.10.10.1 architecture.}
\label{fig:annz}
\end{figure}

For FSC sources with 2MASS counterparts, we choose the elliptical isophotal aperture based on the $K_s$ 20 mag/arcsec$^2$ isophote which gives a good estimate of the integrated flux and colour. The 2MASS training set contains 21050 galaxies, 6664 of which are separated out to form a validation set. The 2MASS testing set contains 6771 galaxies. For FSC sources with SDSS counterparts, the SDSS model magnitudes obtained through the best-fit model (either a pure deVaucouleurs or an exponential) in the r band is a good choice for accurate colours. The SDSS training set contains 10096 galaxies, the testing set contains 5586 galaxies and the validation set contains 1856 galaxies. Extinction corrected is applied using the full-sky 100\,$\mu$m maps (Schlegel, Finkbeiner \& Davis 1998). The colour-colour distribution of the training (black plus signs) and testing set (red plus signs for the 2MASS cross-ids and blue plus signs for the SDSS cross-ids) is shown in Fig.~\ref{fig:colour_dist}. The SDSS training set is not as representative as the 2MASS training set. The photometric redshift error defined as 
\begin{equation}
\sigma=\sqrt{\left\langle \left(\frac{z_{phot}-z_{spec}}{1+z_{spec}}\right)^2 \right\rangle}
\end{equation}
is $1.5\%$ for the 2MASS validation set and $2.3\%$ for the SDSS validation` set (see Fig.~\ref{fig:annz}). 

\subsection{THE TEMPLATE-FITTING TECHNIQUE}

Another widely-used photometric redshift estimation technique is the template-fitting method where
the observed fluxes are compared to those from a library of SED templates (based upon observations
or population synthesis models) which represent different galaxy populations. Unlike the training
set method, a representative set of galaxies with known redshifts is not needed. And once the
spectral type and redshift are determined through $\chi^2$ minimisation, we can carry on to
predict fluxes at other wavelengths as well as determining useful parameters like the extinction, 
bolometric luminosity, stellar mass etc.

\begin{figure}
\includegraphics[height=2.8in,width=3.4in]{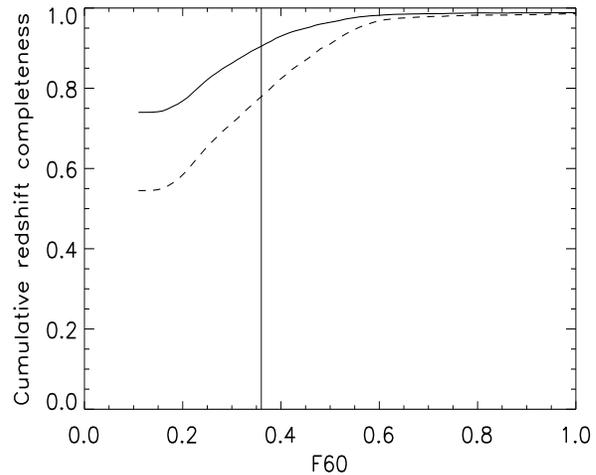}
\includegraphics[height=2.8in,width=3.4in]{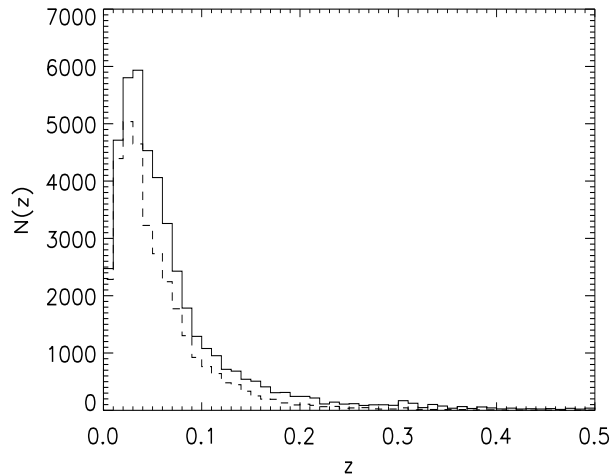}
\caption{Upper: The cumulative redshift completeness as a function of the 60 micron flux density. The dashed line is the completeness curve with only spectroscopic redshifts included and the solid line is with both spectroscopic and photometric redshifts included. The vertical line is S60=0.36 Jy. Lower: Redshift histogram for IIFSCz galaxies.}
\label{fig:redshift_completeness}
\end{figure}

Here we apply the template-fitting method that has been used to construct the SWIRE Photometric
Redshift Catalogue (Rowan-Robinson et al. 2008 and references therein) to the IIFSCz galaxies with either optical or near-infrared photometry. We use only a single pass through the data and use a resolution of 0.002 in $log_{10}(1+z)$. We use 6 galaxy templates (E, Sab, Sbc, Scd, Sdm and starburst) and 3 QSO templates, as in RR08. The resulting rms error in $(1+z)$ for galaxies is $2.7\%$ for 2MASS sources (JHK) and $4.4\%$ for SDSS sources (ugriz), and the outlier rate is $<0.1\%$. Although the rms values are slightly worse than the neural network method, the template method does have the advantage that it is able to predict redshifts which lie outside the range of values in the training set. This is a definite issue for the SDSS sources and for these we have adopted the order of priority (1) spectroscopic redshift, if available, (2) template method redshift, (3) neural network method redshift. For other sources (2MASS sources or NVSS sources), we have adopted (1) spectroscopic redshift, (2) neural network method redshift, (3) template method redshift.

We are able to determine a redshift for around 74$\%$ of the sources in our catalogue. For these sources we use the infrared template fitting method of RR08 to fit four infrared templates (cirrus, M82 starburst, Arp220 starburst, AGN dust torus) to the 12-100 $\mu$m IRAS data, using spectroscopic redshifts where available.  Parameters of these infrared template fits are given in the catalogue, including the infrared luminosities in each component.  These template fits are used to predict the fluxes at 12, 25, 60, 90, 100, 110, 140, 160, 250, 350, 500, 850, 1250, and 1380 $\mu$m, which cover the survey wavelengths of AKARI, \begin{em}Planck\end{em}-Surveyor and \begin{em}Herschel\end{em}. We do not attempt infrared template fits for sources with $z  < 0.0003$ (essentially Local Group objects), since the FSS fluxes for these are likely to be serious underestimates. We believe the combination of precise optical, radio or near infrared positions, redshifts and predicted submillimetre fluxes will make this catalogue invaluable for future large-area far infrared and submillimetre surveys.

For the remaining $\sim$15,000 sources we do not at the moment have optical identifications or good positions, but we note that most of those that fall outside the areas surveyed by SDSS will be similar to the SDSS sources.  

\section{Catalogue Descriptions}
\label{sec:catalogueDescriptions}
The columns in the IIFSCz include source name (as appears in the IRAS FSC), position and flag (1=SDSS, 2=2MASS, 3=NVSS, 4=NED, 5=FSC and prioritised in the same order), IRAS fluxes and flux-quality flags, $ugrizJHK$ magnitudes and errors, photometry flag (1=2MASS XSC, 2=2MASS PSC), integrated 1.4 GHz flux density and error, spectroscopic redshift and flag (1=SDSS, 2=PSCz, 3=FSSz, 4=6dF, 5=NED and prioritised in the same order), $ANNz$ photometric redshift and error, template-fitting photometric redshift, adopted redshift, optical galaxy template type, extinction Av from optical galaxy template fit, reduced $\chi^2$ for galaxy template fit, absolute B magnitude, optical bolometric luminosity, fraction of contribution at 60\,$\mu$m of cirrus, M82, AGN and A220 infrared template, bolometric luminosity in cirrus, M82, AGN and A220 component, infrared luminosity, infrared template type (=1 for cirrus galaxies, =2 for M82 starbursts, =3 for A220 starbursts, =4 for AGN dust tori, i.e. $L_{tor} > L_{M82}$), reduced $\chi^2$ for infrared template fit, predicted fluxes at 12, 25, 60, 90, 100, 110, 140, 160, 250, 350, 500, 850, 1250 and 1380\,$\mu$m, other source names and types. The Catalogue and description are available from http://astro.imperial.ac.uk/$\sim$mrr/fss/. 

\begin{figure*}
\includegraphics[height=3.0in,width=5in]{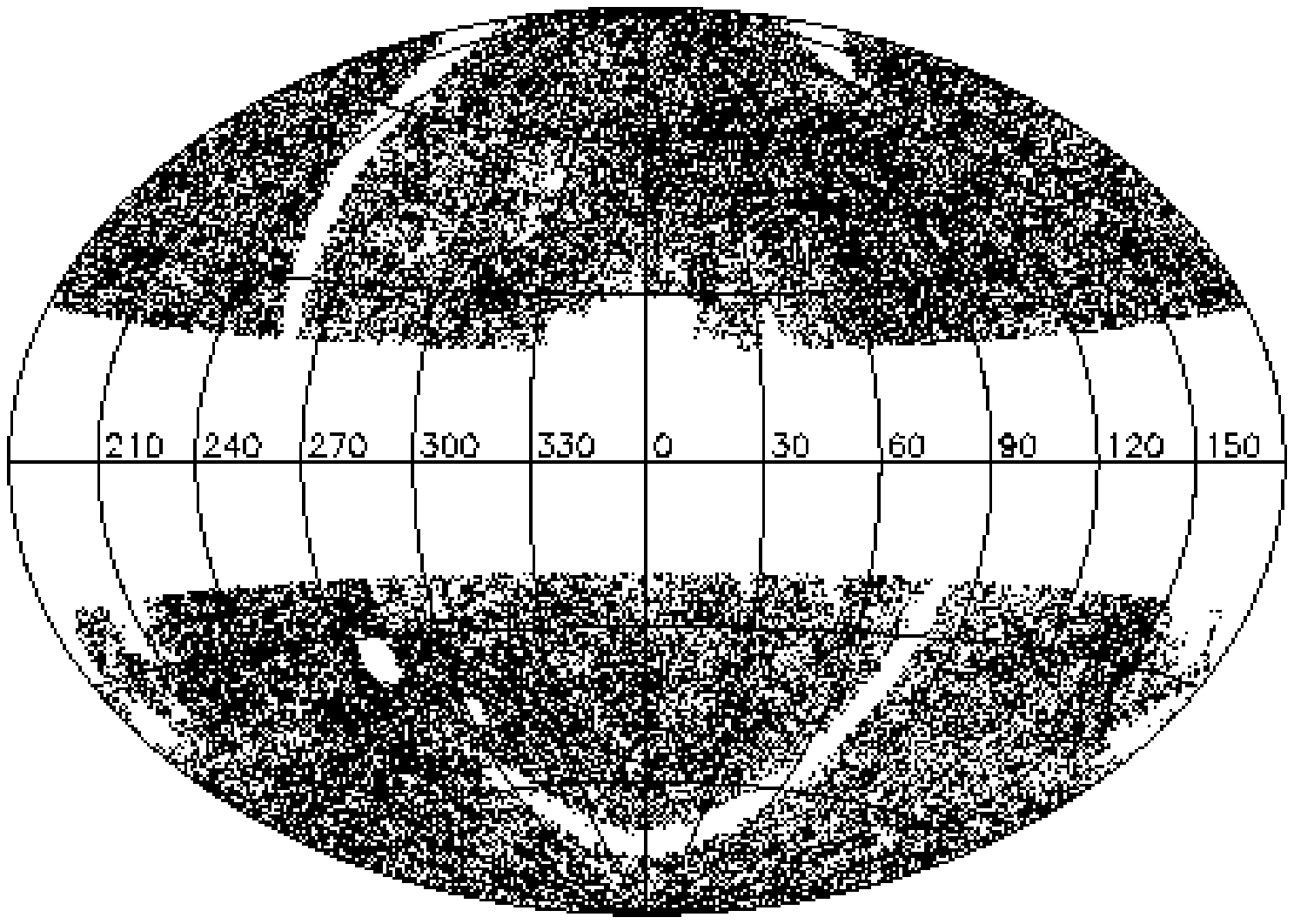}
\includegraphics[height=3.0in,width=5in]{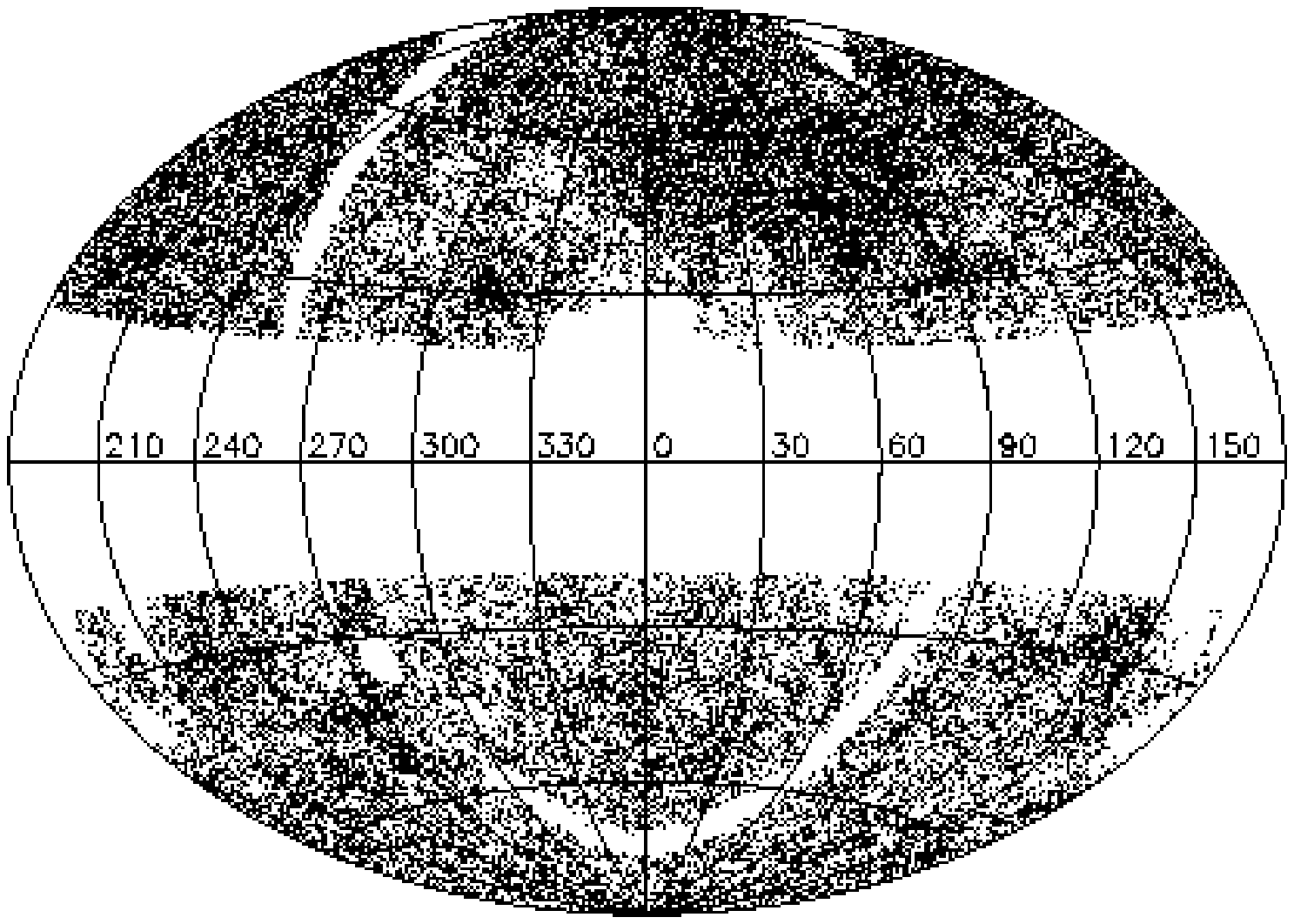}
\includegraphics[height=3.0in,width=5in]{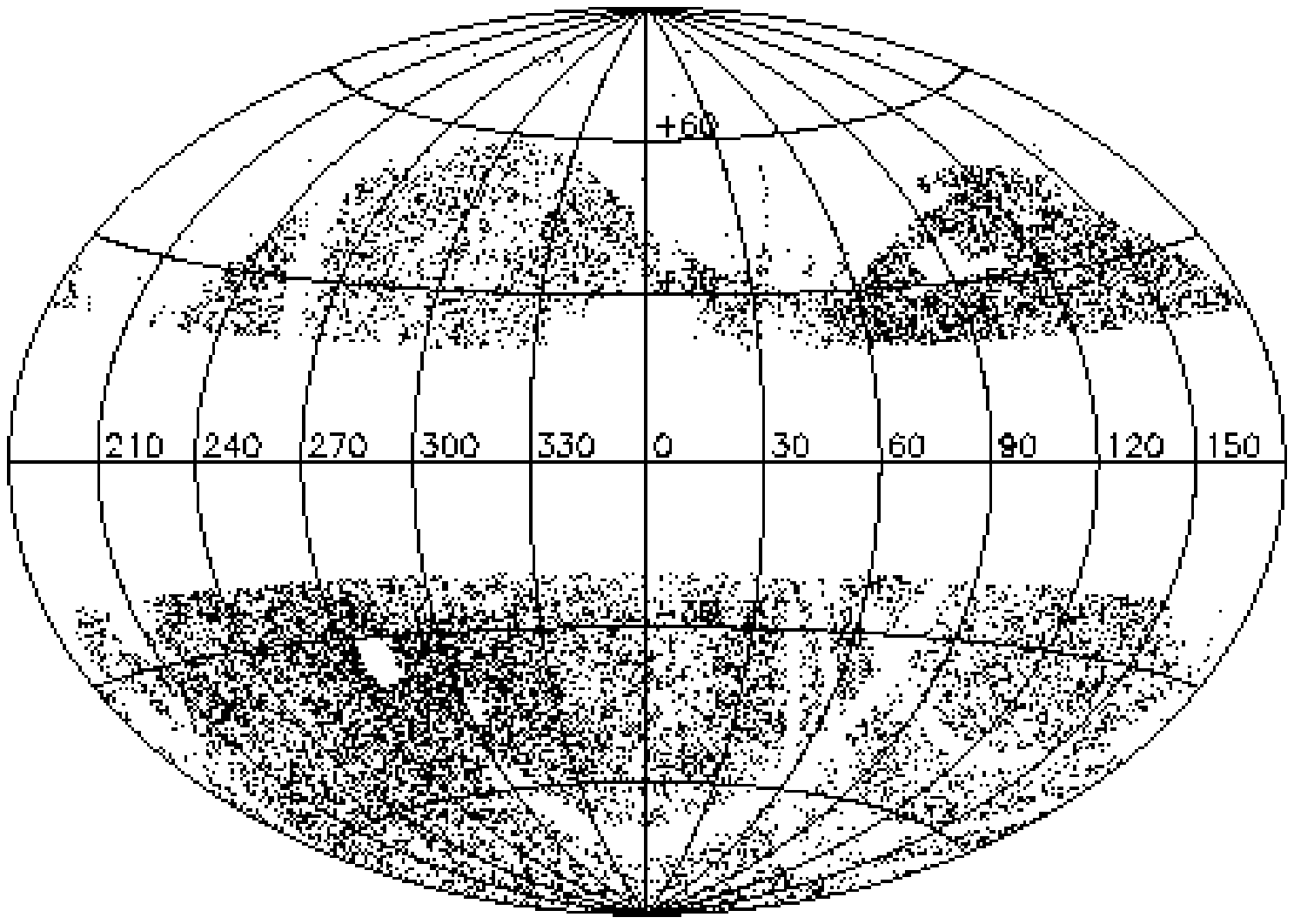}
\caption{The sky distribution of all galaxies in the IIFSCz (top), galaxies with either spectroscopic or photometric redshift (middle) and galaxies without redshift (bottom), in galactic coordinates. The depth of the IIFSCz is affected by variations in the IRAS coverage across the sky. Note that in the bottom panel, the redshift completeness is very high in regions covered by the SDSS DR6 survey.}
\label{fig:sky_dist}
\end{figure*}

In Fig.~\ref{fig:redshift_completeness}, the cumulative redshift completeness is plotted against the 60\,$\mu$m flux. The vertical line shows the intrinsic $90\%$ completeness limit of the FSC, 0.36 Jy at 60\,$\mu$m. Consequently, for research programmes such as the cosmological dipole where sample completeness is desired, the upper panel in Fig. 7 should be taken into account. The IIFSCz covers about $61\%$ of the whole sky. In Fig.~\ref{fig:sky_dist}, the sky distribution is plotted for all galaxies in the IIFSCz, galaxies with either spectroscopic or photometric redshift and galaxies without any redshift estimate. 

To summarise, the IIFSCz contains a total of 60,303 galaxies, $55\%$ of which have spectroscopic redshifts from NED, FSSz, PSCz, 6dF and the SDSS spectroscopic DR6 survey (see the breakdown of spectroscopic redshift sources in Table~\ref{zSource}) and $20\%$ of which have photometric redshifts from either the empirical training set or the template-fitting method. At a flux limit of S60=0.36 Jy, more than $90\%$ of the galaxies in the IIFSCz have either spectroscopic or photometric redshifts. There are 344 known QSOs and only 6 of these do not have redshift information.

\begin{table}
\caption{Source of spectroscopic redshift for the IIFSCz}\label{zSource}
\begin{tabular}[pos]{lll}
\hline
Source of $z_{spec}$         & Number   & Fraction \\
\hline
NED                          & 23,703    &$39.3\%$\\
IRAS FSSz                    & 568      &$0.9\%$\\
IRAS PSCz                    & 1,972     &$3.3\%$\\
6dF Galaxy Survey            & 2,794     &$4.6\%$\\
SDSS spectroscopic DR6       & 3,844     &$6.4\%$\\
\hline
Total                        &32,881    &$55.0\%$\\
\hline
\end{tabular}

\end{table}

\section{DISCUSSIONS}
\label{sec:discussions}

\begin{figure}
\includegraphics[height=3.3in,width=3.4in]{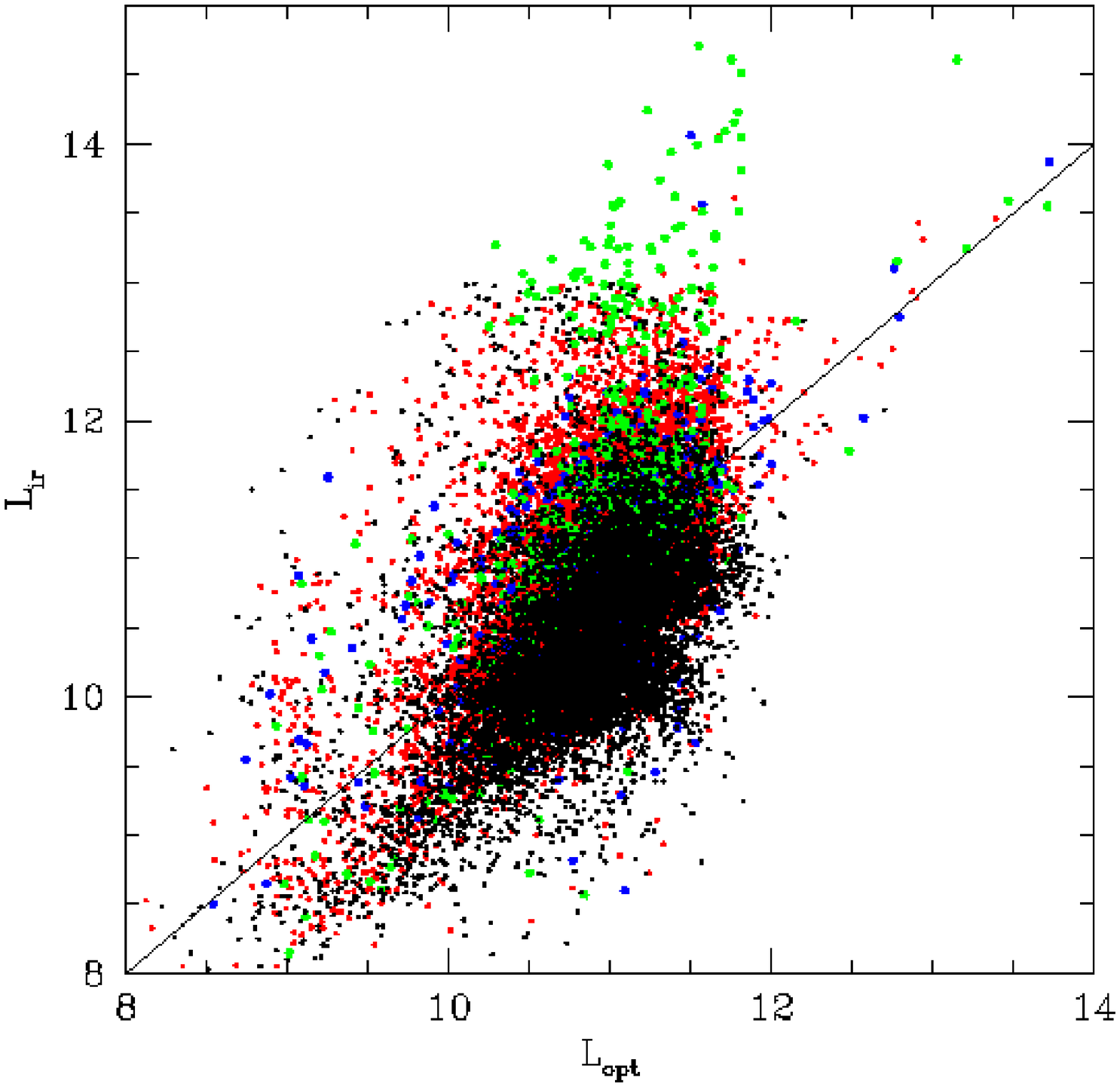}
\caption{Infrared bolometric luminosity ($L_{ir}$) versus optical bolometric luminosity ($L_{opt}$), colour-coded by the infrared template type.}
\label{fig:luminosity}
\end{figure}

Fig.~\ref{fig:luminosity} shows the distribution of infrared bolometric luminosity, $L_{ir}$, versus optical bolometric luminosity, $L_{opt}$, from our template fitting. The figure is colour-coded by the infrared template type, and only galaxies with at least two detected far infrared fluxes are included. Most cirrus galaxies have $L_{ir} < L_{opt}$, as expected if the emission is from an optically thin interstellar dust distribution. However there are some cirrus galaxies with $L_{ir} > L_{opt}$, indicating a higher dust optical depth. There is also the interesting population of cool luminous galaxies, with $L_{ir} > 10^{12} L_{\odot}$, discussed by Rowan-Robinson et al. (2005, 2008). The highest infrared luminosities are dominated by Arp 220 template types, but there are significant numbers of ultraluminous infrared galaxies which are M82 template types. Objects with $L_{opt} > 10^{12} L_{\odot}$ are Type 1 QSOs and most have $L_{ir} < L_{opt}$, as expected if the infrared emission is dominated by a dust torus illuminated by the QSO.

\begin{figure}
\includegraphics[height=3.3in,width=3.4in]{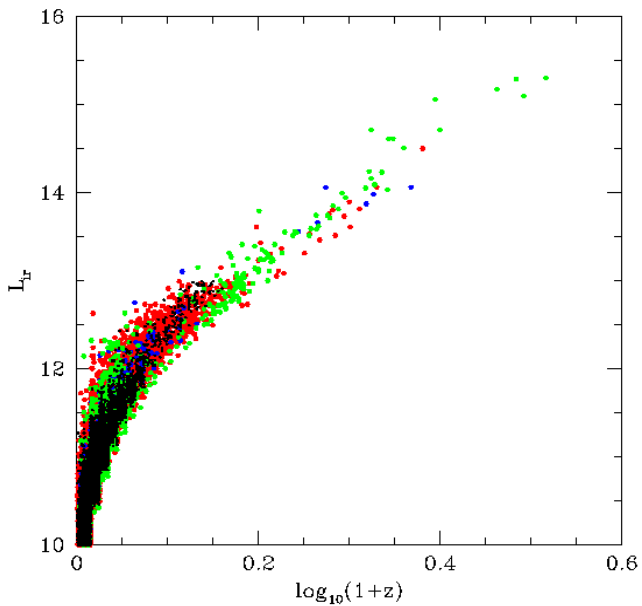}
\caption{Infrared bolometric luminosity versus redshift, colour-coded by the infrared template type.}
\label{fig:luminosity-redshift}
\end{figure}

One of the interesting discoveries of the IRAS Faint Source Catalog was the existence of hyperluminous infrared galaxies (Rowan-Robinson et al. 1991, Rowan-Robinson 2001), galaxies with infrared luminosities $> 10^{13} L_{\odot}$. In our Catalogue we now find 159 hyperluminous infrared galaxies with S60 $>$ 0.2 Jy, of which 38 have S60 $>$ 0.36 Jy, our completeness limit.  Fig.~\ref{fig:luminosity-redshift} shows the distribution of bolometric infrared luminosity, $L_{ir}$, versus redshift, colour-coded by the infrared template type. Only galaxies with more than one far infrared flux detected are included. In our template fitting we do not permit the cirrus template to be used if $L_{ir} > 10^{13} L_{\odot}$. We see that the highest luminosities are dominated by the high optical depth starburst, Arp 220, templates, but with some sources dominated by AGN dust tori and a few M82 starbursts.

\begin{figure}
\includegraphics[height=3.3in,width=3.4in]{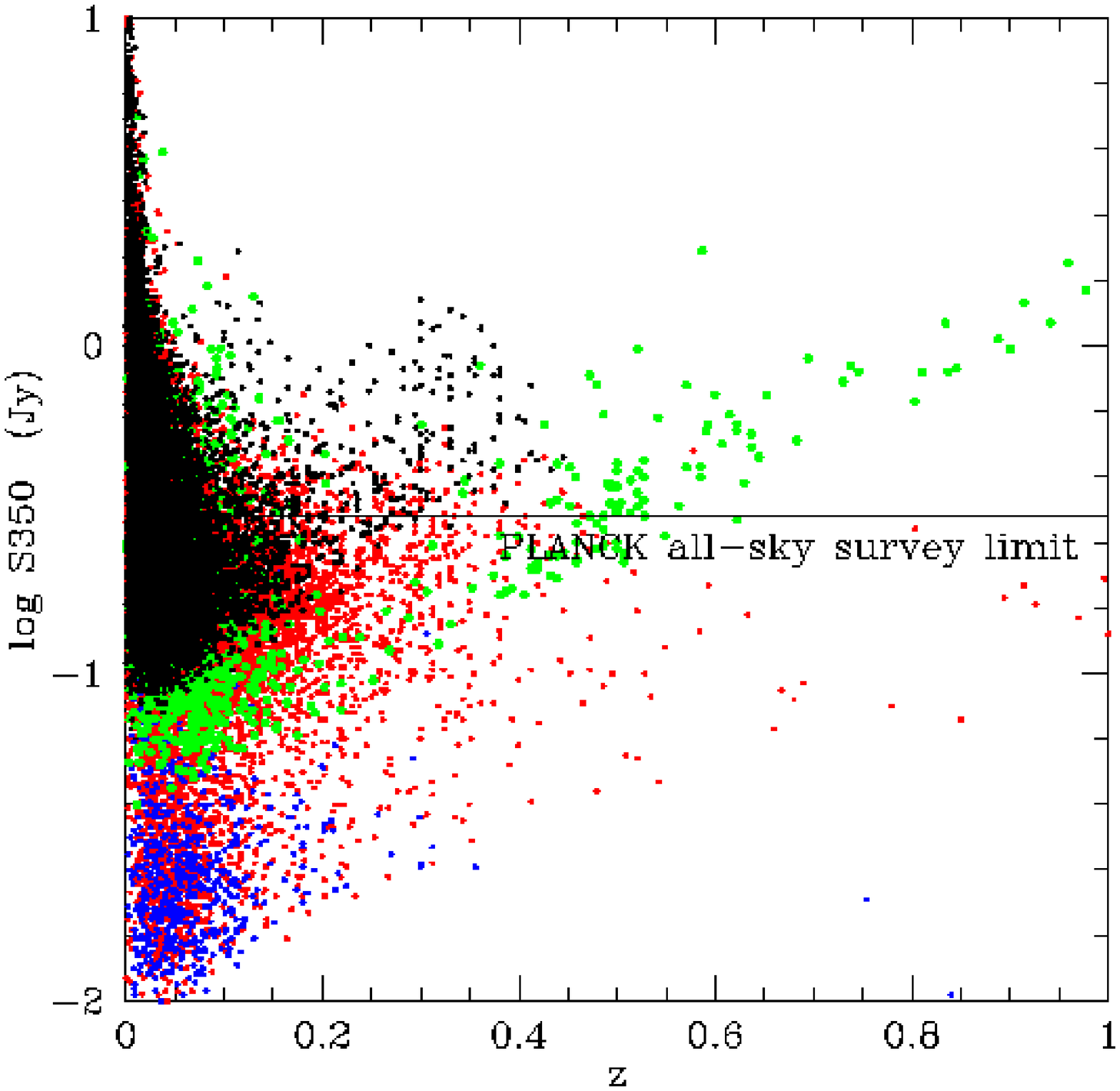}
\caption{The predicted 350\,$\mu$m flux in Jy versus redshift, colour-coded by the infrared template type.}
\label{fig:S350}
\end{figure}

There are 50 galaxies with $L_{ir} > 10^{14} L_{\odot}$, of which 15 are QSOs, and 8 with $L_{ir} > 10^{15} L_{\odot}$. The latter include the well known hyperluminous galaxies IRAS F10214+4724 and 08279+5255, both of which are lensed and are discussed in Rowan-Robinson (2001). Five of these eight galaxies have spectroscopic redshifts. The hyperluminous infrared galaxies in our Catalogue will be discussed in a subsequent paper (Rowan-Robinson et al. 2009, in preparation).

Fig.~\ref{fig:S350} illustrates our predicted submillimetre fluxes derived from our template fits. It shows the predicted 350 $\mu$m flux versus redshift, again colour-coded by infrared template type. We have indicated the predicted flux limit of the \begin{em}Planck\end{em} Surveyor all-sky survey. We are predicting that over 23000 of the sources in our Catalogue will have 350 $\mu$m fluxes above the \begin{em}Planck\end{em} all-sky survey limit of 0.3 Jy. \begin{em}Planck\end{em} should detect significantly more sources than this because (i) not all the sources in our Catalogue have redshift estimates and template fits, (ii) some high redshift infrared galaxies may fall below our 60 $\mu$m detection limit but still be detectable at 350 $\mu$m. Our Catalogue should however include all sources detected by \begin{em}Planck\end{em} with z $<$ 0.5.

\section{CONCLUSIONS}
\label{sec:conclusions}

We have presented the Imperial IRAS-FSC Redshift Catalogue (IIFSCz). It contains 60,303 galaxies selected at 60 $\mu$m from the IRAS Faint Source Catalog, covering around $61\%$ of the whole sky. The process of retrieving spectroscopic redshifts, multiwavelength cross-matching and photometric redshift estimation is described in some detail. In the Catalogue, we give the best possible position, IRAS fluxes, optical, near-infrared and/or radio identifications, spectroscopic redshift (if available) or photometric redshift (if possible), predicted fluxes at wavelengths ranging from 12 to 1380 $\mu$m. Overall, 32,881 galaxies in the IIFSCz (55$\%$) have received spectroscopic redshifts from past redshift surveys such as the IRAS PSCz, FSSz and 6dF and around 12,000 galaxies ($\sim20\%$) obtained photometric redshifts through either the training set (for sources with 2MASS or NVSS photometry) or the template-fitting method (for sources with SDSS photometry). At a flux limit of S(60)=0.36 Jy, the redshift completeness of the Catalogue, including both spectroscopic and photometric redshifts, is increased to 90$\%$.

The IIFSCz provides a huge data set for large-scale structure studies and validations of recent and future infrared and submillimetre surveys (e.g. AKARI, \begin{em}Planck\end{em} and \begin{em}Herschel\end{em}). Potential users should be aware of issues such as the intrinsic IRAS FSC completeness limit, the redshift completeness variations across the sky and the varying quality of photometric redshift derived for different subsets of the Catalogue.

\section*{ACKNOWLEDGEMENTS}
We thank Seb Oliver for providing the FSSz catalogue and helpful discussions on obtaining optical identifications in the SDSS DR6 survey. This research has made extensive use of the NASA/IPAC EXTRAGALACTIC DATABASE (NED) and the Sloan Digital Sky Survey (SDSS). L.W. thanks Marion Schmitz and Joseph Mazzarella for their help on various catalogues in the NED. L.W. is supported by a Dorothy Hodgkin Postgraduate Award (DHPA).

The NASA/IPAC EXTRAGALACTIC DATABASE (NED) is operated by the JET PROPULSION LABORATORY, CALTECH, under contract with the NATIONAL AERONAUTICS AND SPACE ADMINISTRATION.

Funding for the Sloan Digital Sky Survey (SDSS) and SDSS-II has been provided by the Alfred P. Sloan Foundation, the Participating Institutions, the National Science Foundation, the U.S. Department of Energy, the National Aeronautics and Space Administration, the Japanese Monbukagakusho, and the Max Planck Society, and the Higher Education Funding Council for England. The SDSS Web site is http://www.sdss.org/.

The SDSS is managed by the Astrophysical Research Consortium (ARC) for the Participating Institutions. The Participating Institutions are the American Museum of Natural History, Astrophysical Institute Potsdam, University of Basel, University of Cambridge, Case Western Reserve University, The University of Chicago, Drexel University, Fermilab, the Institute for Advanced Study, the Japan Participation Group, The Johns Hopkins University, the Joint Institute for Nuclear Astrophysics, the Kavli Institute for Particle Astrophysics and Cosmology, the Korean Scientist Group, the Chinese Academy of Sciences (LAMOST), Los Alamos National Laboratory, the Max-Planck-Institute for Astronomy (MPIA), the Max-Planck-Institute for Astrophysics (MPA), New Mexico State University, Ohio State University, University of Pittsburgh, University of Portsmouth, Princeton University, the United States Naval Observatory, and the University of Washington.


\begin{thebibliography}{99}
\bibitem{} Appleton P.N. et al., 2004, ApJSS, 154, 147
\bibitem{} Becker R.H., White R.L., Helfand D.J., 1995, AJ, 450, 559
\bibitem{} Brusa et al., 2007, ApJS, 172, 353 
\bibitem{} Ciliegi P., Zamorani G., Hasinger G., Lehmann I., Szokoly G., Wilson G., 2003, AA, 398, 901
\bibitem{} Collister A.A., Lahav O., 2004, PASP, 116, 345
\bibitem{} Condon J.J, Cotton W.D., Greisen E.W., Yin Q.F., Perley R.A., Taylor G.B., Broderick J.J., 1998, AJ, 115, 1693
\bibitem{} de Jong T., Klein U., Wielebinski R., Wunderlich E., 1985, AA, 147, L6
\bibitem{} Eisenstein et al., 2005, ApJ, 633, 560
\bibitem{} Erdo\v{g}du P. et al., 2006, MNRAS, 368, 1515
\bibitem{} Firth A.E., Lahav O., Somerville R.S., 2003, MNRAS, 339, 1195 
\bibitem{} Griffin et al., 2007, AdSpR, 40, 612
\bibitem{} Hawkins E., Maddox, S., Branchini E., Saunder W., 2001, MNRAS, 325, 589
\bibitem{} Helou G., Soifer B.T., Rowan-Robinson M., 1985, ApJ, 298, L7
\bibitem{} Jarrett T.H., Chester T., Cutri R., Schneider S.E., Huchra J.P., 2003, AJ, 125, 525
\bibitem{} Jones et al., 2004, MNRAS, 355, 747
\bibitem{} Jones et al., 2005, PASA, 22, 277
\bibitem{} Kawada M. et al., 2007, PASJ, 59, S389
\bibitem{} Maller et al., 2003, ApJ, 598, L1
\bibitem{} Mann R.G., Saunders W., Taylor A.N., 1996, MNRAS, 279, 636
\bibitem{} Moshir M. et al., 1992, Explanatory Supplement to the IRAS Faint Source Survey, Version 2, JPL D-10015 8/92 (Pasadena:JPL)
\bibitem{} Murakami H. et al., 2007, PASJ 59, S369
\bibitem{} Pilbratt G., 2004, The Herschel mission: status and observing opportunities. Proc. SPIRE 5487, 401
\bibitem{} Rice W., Lonsdale C.J., Soifer B.T., Neugebauer G., Kopan E.L., Lloyd L.A., de Jong T., Habing H.J., 1988, ApJS, 68, 91 
\bibitem{} Rowan-Robinson M. et al., 1991, Nat., 351, 719
\bibitem{} Rowan-Robinson M. et al., 2000, MNRAS, 314, 375
\bibitem{} Rowan-Robinson M., 2001, IAUS, 204, 265
\bibitem{} Rowan-Robinson M. et al., 2004, MNRAS, 351, 1290
\bibitem{} Rowan-Robinson M. et al., 2005, AJ, 129, 1183 
\bibitem{} Rowan-Robinson M. et al., 2008, MNRAS, 386, 697
\bibitem{} Saunders W. et al., 2000, MNRAS, 317, 55
\bibitem{} Serjeant S., Harrison D., 2005, MNRAS, 356, 192
\bibitem{} Schlegel D.J., Finkbeiner D.P., Davis M., 1998, ApJ, 500, 525
\bibitem{} Skrutskie M.F. et al., 1997, in ASSL Vol. 210, The Impact of Large Scale Near-IR Sky Surveys, ed. F. Garz\'{o}n et al. (Dordrecht:Kluwer), 25
\bibitem{} Sutherland W., Saunders W., 1992, MNRAS, 259, 413
\bibitem{} Szapudi I., Branchini E., Frenk C.S., Maddox S., Saunders W., 2000, MNRAS, 318, L45
\bibitem{} Wang L. et al., 2008, MNRAS, 387, 601
\bibitem{} Wolstencroft R.D., Savage A., Clowes R.G., MacGillivray H.T., Leggett S.K., Kalafi M., 1986, MNRAS, 223, 279
\bibitem{} York D.G. et al., 2000, AJ, 120, 1579
\bibitem{} Yun M.S., Reddy N.A., Condon J.J., 2001, ApJ, 554, 803
\end{thebibliography}
\end{document}